# Parity-time symmetric optical neural networks

HAOQIN DENG,[1] MERCEDEH KHAJAVIKHAN[1,*]

[1]*Ming Hsieh Department of Electrical and Computer Engineering, University of Southern California, Los Angeles, California 90089, USA*
*Corresponding author: khajavik@usc.edu*

**Optical neural networks (ONNs), implemented on an array of cascaded Mach-Zehnder interferometers (MZIs), have recently been proposed as a possible replacement for conventional deep learning hardware. They potentially offer higher energy efficiency and computational speed when compared to their electronic counterparts. By utilizing tunable phase shifters, one can adjust the output of each of MZIs in order to enable emulation of arbitrary matrix-vector multiplication. These phase shifters are central to the programmability of ONNs, but they require large footprint and are relatively slow. Here we propose an ONN architecture that utilizes parity-time (PT) symmetric couplers as its building blocks. Instead of modulating phase, gain/loss contrasts across the array are adjusted as a means to train the network. We demonstrate that PT symmetric optical neural networks (PT-ONN) are adequately expressive by performing the digit-recognition task on the modified national institute of standard and technology (MNIST) dataset. Compared to conventional ONNs, the PT-ONN achieves a comparable accuracy (67% vs. 71%) while circumventing the problems associated with changing phase. Our approach may lead to new and alternative avenues for fast training in chip-scale optical neural networks.**

## 1. Introduction

The computing power of modern electronics, which adopt the Von-Neumann architecture, is inherently bottlenecked by the data transfer rate between the processing and memory units. Emerging computing architectures, such as neuromorphic approaches [1,2], represent more effective computational schemes by intertwining logic with memory. In recent years, optical platforms have once again been proposed as a promising candidate for fully/partially replacing the electronic-based computing machines. Optical computing is particularly of interest because of the prospect of requiring lower energy per bit and having less latency [3–10]. In 2017, a team of researchers from MIT demonstrated a ground-breaking, fully integrated optical neural network on a silicon chip [3] by cascading a number of Mach-Zehnder interferometers (MZIs). An arbitrary matrix can be effectively mapped onto this ONN hardware by computing the corresponding phases of each MZI. For such networks, the required nonlinearities can be implemented through various approaches that utilize intensity modulators [11], the saturation effect of cameras [12], quadratic nonlinearity of photodiodes [13], saturation of semiconductor amplifiers [14], and saturable absorbers [15–17], to name a few. Since then, a number of schemes have been proposed to further optimize the implementation of these arrays and their on-chip training processes [18–21].

While optical neural networks are receiving considerable attention in both academic and industrial settings, it is now clear that changing phases on chip is undesirable and can significantly overshadow the potential benefits of the photonic accelerators [22,23]. In these arrangements, phase changing is typically accomplished by thermo-optical phase shifters, where a bias current is applied to change the refractive index of an optical waveguide through the thermo-optic effect [3,24]. However, since the thermo-optic coefficient of most optoelectronic materials is relatively small, translating it to a phase change requires a path length that is typically on the order of tens to hundreds of micrometers [24]. Given that for processing $N$ bits of data, $O(N^2)$ phase shifters are needed, such schemes can lead to prohibitively large structures as the size of the data increases. Moreover, the time it takes for the phase change to take effect is relatively long, on the order of tens of microseconds [24], which can limit the speed of on-chip training processes, where one needs to frequently vary phases to compute gradients. A number of recent works have aimed to address these problems by proposing alternative architectures that make use of optical fast Fourier transform (OFFT) [23], ring resonators [25,26], acousto-optic modulators [27], and 3D printing [22]. Other approaches based on phase-change materials, electro-absorption and electrooptic effect may also solve some of these issues, but the technology is still maturing [28–31].

However, the choice of cascaded passive MZIs for implementing ONNs is not related to the fundamentals of neural networks; rather, it comes from the mathematical convenience of expressing an arbitrary matrix into MZI-representable sub-systems through unitary matrices and singular value decomposition (SVD) [32,33]. It is well known that such unitary matrices can be readily implemented in passive optical platforms like silicon or silicon nitride wafers using a combination of MZIs. Nevertheless, since the original matrix ($W_{i,j}$) is generally non-unitary, amplification/attenuation has to inevitably be deployed in the optical implementation of ONNs. In addition, laser light is already used in such networks. With on-chip optical settings, lasing is typically achieved by pumping and carrier injection in appropriate III-V compound semiconductors. Moreover, saturable absorbers are considered as one of the choices for activation function in the optical domain [3,17]. Most such elements are based on III-V semiconductors as well. Finally, as the network becomes larger, optical amplification may be needed in order to compensate the inevitable optical losses. Given the omnipresence of amplification in optical neural networks, it may be beneficial to explore alternative ONN architectures in which gain/loss is used in lieu of phase shifters.

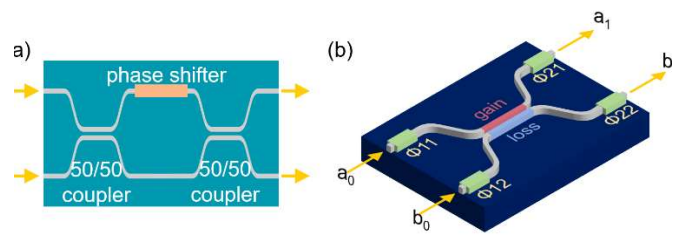

**Fig. 1.** (a) A Mach-Zehnder interferometer composed of two cascaded 50/50 beam splitters, with a phase shifter sandwiched in between. (b) A PT-symmetric directional coupler comprising of a pair of waveguides, one experiencing gain and the other one a similar amount of loss. Constant phases ($\phi_{11}, \phi_{12}, \phi_{21}, \phi_{22}$) are added in the input and output ports to turn the transfer matrix entirely real.

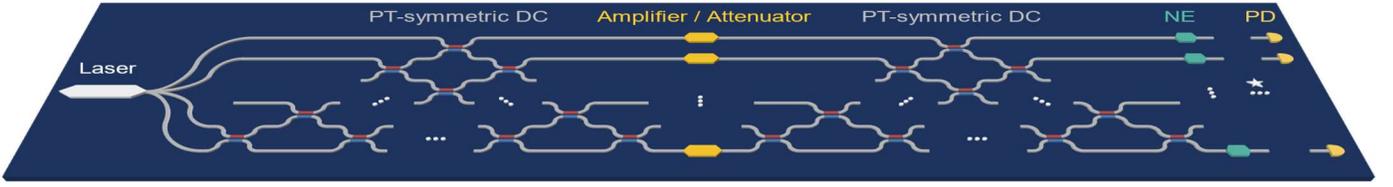

**Fig. 2**. The overall structure of a 2-layer PT-ONN: In layer one, lasers encode $N_1$ pixels. The optical signals are first sent to a triangular-shaped array of $(N_1(N_1-1)/2)$ PT-symmetric couplers. The light then passes through $N_2$ amplifiers/attenuators followed by the second triangular-shaped array of $(N_2(N_2-1)/2)$ PT-couplers before encountering $N_2$ nonlinear elements. The second layer, represented by the star, contains similar elements, but with $N_2$ and $N_3$ values. This layer terminates with $N_3$ photo detectors. The values $N_1$, $N_2$, $N_3$ represent the dimensionality of input, hidden, and output layers, respectively. DC: Directional Coupler; NE: Nonlinear Element; PD: Photodetector.

In this paper, we propose a new architecture based on parity-time (PT) symmetric couplers [34] that can partially address some of the problems of current ONNs by using optical gain/loss in III-V semiconductors or other gain materials. We refer to this architecture as parity-time symmetric optical neural network (PT-ONN). It borrows the cascading structure from [3] to ensure that a large number of free parameters are available and that the network is sufficiently expressive to distinguish patterns. We show that even at low/moderate levels of gain/loss contrast, our network can provide a comparable performance to that of passive optical systems with phase shifters. Some practical considerations concerning the physical realization of these networks will also be discussed. As will be shown, our approach of replacing phase shifters with PT-symmetric couplers has the potential to significantly reduce the energy consumption, increase the training speed, and lower the footprint in on-chip optical neural networks. More novel and practical PT configurations can be used to further improve the operation of the optical neural networks.

## 2. PT-ONN architecture

The main building block of PT-ONN is a two-level parity-time symmetric directional coupler whose gain/loss factors can be tuned either individually or together [34]. In general, a structure is considered to be PT-symmetric if it is invariant under the simultaneous action of the P (space) and T (time) inversion operators. Despite having a non-Hermitian representation, these systems may still support entirely real spectra (eigenvalues). While originally developed in the context of quantum mechanics, PT-symmetric notions have lately attracted considerable attention in different areas of optics, including photonic lattices, micro resonators, gratings, sensors, wireless power transfer, and lasers, to name a few [35–41]. In optical settings, a structure is PT symmetric if the real part of the refractive index is an even function of space, while the imaginary component (representing gain and loss) exhibits an odd profile. Here, a PT-coupler refers to a coupled waveguide system in which one channel experiences gain and the other one an equal amount of loss. Consequently, the propagation constants are the eigenvalues, and the electromagnetic modes represent the eigenvectors of the system. The ratio of gain-loss contrast to coupling serves as a parameter that largely determines the response of the structure. In fact, when this ratio becomes equal to unity, it can be shown that both eigenvalues and eigenvectors of the structure coalesce. This point that represents a spontaneous symmetry breaking is known as an exceptional point. In this study, we operate our PT couplers in the PT unbroken regime, where the governing parameter is less than unity and the system works below the exceptional point [35]. Figure 1 compares the PT-coupler with a tunable MZI system.

In a PT-coupler, the energy exchange between the two waveguides obeys the following system of equations [34]:

$$\begin{cases} i\frac{da}{dz} - i\frac{g}{2}a + \kappa b = 0 \\ i\frac{db}{dz} + i\frac{g}{2}b + \kappa a = 0 \end{cases} \quad (1)$$

where $a$ and $b$ represent the electric field in the two waveguides, $\kappa$ is the coupling strength, $z$ the propagation length, and $g$ signifies the gain/loss contrast. The relationship between the input ($a_0$ and $b_0$) and output ($a$ and $b$) ports (see Fig. 1b) can be derived in two different regimes of operation. Below the PT symmetry breaking point (where $g/2\kappa \leq 1$), the coupling matrix is expressed by:

$$\begin{bmatrix} a \\ b \end{bmatrix} = \frac{1}{\cos\theta}\begin{bmatrix} \cos(Z\cos\theta - \theta) & i\sin(Z\cos\theta) \\ i\sin(Z\cos\theta) & \cos(Z\cos\theta + \theta) \end{bmatrix}\begin{bmatrix} a_0 \\ b_0 \end{bmatrix}, \quad (2)$$

where $\theta = \sin^{-1}(g/2\kappa)$ and $Z = \kappa z$. On the other hand, in the PT-broken phase (i.e., above the PT-symmetry breaking point) the PT-coupler behaves according to:

$$\begin{bmatrix} a \\ b \end{bmatrix} = \frac{1}{\sinh\eta}\begin{bmatrix} \sinh(Z\sinh\eta + \eta) & i\sinh(Z\sinh\eta) \\ i\sinh(Z\sinh\eta) & \sinh(\eta - Z\sinh\eta) \end{bmatrix}\begin{bmatrix} a_0 \\ b_0 \end{bmatrix}, \quad (3)$$

where $g/2\kappa = \cosh\eta$. As expected, in both scenarios, the transfer matrices are non-unitary, due to the inherent non-Hermiticity of the device.

In this work we use PT- couplers exclusively in the PT-unbroken phase. In other words, the gain-loss contrast in the system is only minimally perturbed around zero values (here $g/2\kappa < 0.2$). By adding appropriate constant phases to the input ($-\pi/2$, $-\pi$) and output arms ($\pi/2$, $\pi$), respectively, the transfer function can be modified to only act in real space:

$$\begin{bmatrix} a \\ b \end{bmatrix} = \frac{1}{\cos\theta}\begin{bmatrix} \cos(Z\cos\theta - \theta) & -\sin(Z\cos\theta) \\ \sin(Z\cos\theta) & \cos(Z\cos\theta + \theta) \end{bmatrix}\begin{bmatrix} a_0 \\ b_0 \end{bmatrix}. \quad (4)$$

In our network, we also assume a constant $\kappa$ and $z$ for all couplers, where $Z = \kappa z = 1$. This leaves us with the gain-loss contrasts ($g$'s) as the only on-chip parameters to be used for training (i.e. no phase modulation is required). This can be readily achieved in standard III-V semiconductor systems by pumping/carrier injection. Since varying gain/loss coefficients can be more efficient than changing phases in terms of space, power consumption, and speed, our PT-ONN architecture can potentially require a smaller footprint and accelerate on-chip training at lower powers.

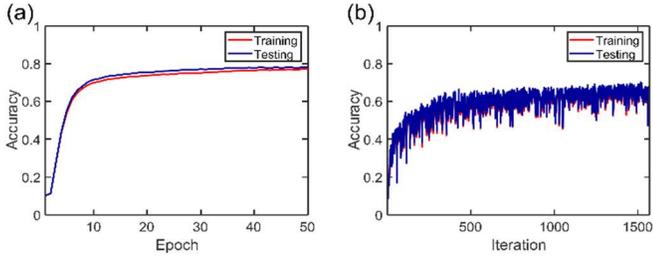

**Fig. 3.** (a) Training and testing accuracies of a classic neural network using backpropagation method. (b) Training and testing accuracies of an optical neural network made of MZIs and using phase shifters as parameters.

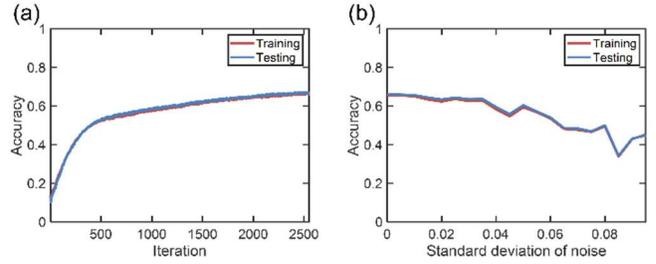

**Fig. 4.** (a) Training and testing accuracies of the parity-time-symmetric neural network. (b) Training and testing accuracies of PT-ONN in the presence of noise. Here the accuracies are normalized against the zero-noise situation.

## 3. Computational experiments and results

Figure 2 shows a schematic of a two-layer PT-symmetric optical neural network that is used in our simulations. In layer 1, $N_1$ pixels of the incoming data are encoded in light amplitude (provided by a series of laser sources/beams). After modulating the data on the carrier frequency, it travels in a triangular-shaped array containing $N_1(N_1 - 1)/2$ PT-symmetric couplers, followed by $N_2$ amplifiers/attenuators, another triangular-shaped array of PT-couplers containing $N_2(N_2 - 1)/2$ components, and finally $N_2$ nonlinear elements. Layer 1 is followed by layer 2, which is similar to the first layer in architecture but with different number of elements ($N_2$ and $N_3$ instead of $N_1$ and $N_2$) and ends in $N_3$ optical detectors. The output of the detectors is then sent to an electronic circuit to calculate the PT-coupler gain/loss parameters ($\theta$'s) in order to implement the gradient descent algorithm in the training cycles. In this example, $N_1, N_2, N_3$ are the sizes of input, hidden, and output layers, respectively.

The simulations are performed for digit recognition task on MNIST dataset [42]. To accomplish this, the $28 \times 28$-pixel images are subsampled by a factor of 16 to be $7 \times 7$-pixel images for computing efficiency improvement. In our studies, we use the input layer of size $7 \times 7$ ($N_1 = 49$), the hidden layer of size $N_2 = 20$, and an output layer of $N_3 = 10$ dimensionality (corresponding to 10 digits). We use a *sigmoid* activation function for the hidden layer (this choice is regardless of the hardware used for the implementation of the nonlinear function), *SoftMax* activation function for the output layer, and *cross-entropy* as the loss function. The simulations are run with Python programs on an Intel i9-9900k CPU. We also assume that all parameters are randomly initialized. For on-chip training, we compute the numerical gradients of the designed parameters using finite difference method. By forward propagating the network with parameters $\theta_i + \Delta\theta$ and $\theta_i - \Delta\theta$, we can measure the output and compute $f(\theta_i + \Delta\theta)$ and $f(\theta_i - \Delta\theta)$, where $f$ is the loss function that is going to be minimized. We then compute the partial gradient $\partial f/\partial \theta_i = (f(\theta_i + \Delta\theta) - f(\theta_i - \Delta\theta))/2\Delta\theta_i$, and use SGD (Stochastic Gradient Descent) to minimize the loss function.

To allow for appropriate benchmarking, in all the following experiments, we use a 2-layer neural network structure with the same topology, where there are $N_1$ input neurons, $N_2$ hidden neurons, $N_3$ output neurons, and the same set of activation and loss functions. We apply the neural network topology to three experimental settings, with different parameter spaces. First, we simulate a classical neural network with parameters being the weight matrix $W_{ij}$ for each layer. Then, we model an MZI-based optical neural network in which phases of the MZIs serve as the parameters. The MZI mesh is arranged in the triangular fashion inspired by [3], which uses singular-value decomposition (SVD). The schematic of this ONN can be found in the Supplement 1, Section 1. Finally, we replace MZIs with PT couplers. In this case the training parameters are gain/loss factors. We use the same topology of the mesh in the second and third simulations in order to allow a direct comparison to be made.

Using the traditional backpropagation method to compute gradients and the SGD (Stochastic Gradient Descent) method to minimize loss function, we first train the network on the subsampled dataset and achieve a peak training accuracy of 77.5% and a testing accuracy of 78.5% (Fig. 3a). This experiment serves to validate our subsampled image set and the two-layer neural network topology. The reported training and testing accuracies are considered to be the upper-bound for a network of the same topology (topology as in the number of layers, number of neurons in each layer, nonlinearities, and the loss function), since on-chip trainings that operate in different parameter spaces are generally expected to achieve lower accuracies.

Next, we study the optical neural network that emulates the structure used in [3], albeit with different size ($N_1, N_2, N_3$), by simulating the on-chip training process. More specifically, the transfer function between each layer is not represented by a single matrix; rather, it is the product of cascading 2-level transfer-matrices that represent MZIs, where the phases are the parameters to be trained (Please see Supplement 1, Fig. S1). By training the network using the numerical method described above, we achieve a peak training accuracy of 69% and a peak testing accuracy of 70.2% (Fig. 3b).

Finally, we evaluate the performance of the PT-ONN architecture, by choosing $Z$ to be equal to 1. The on-chip training process is the same as above, except that we carry out SGD only on the gain/loss dependent $\theta$ variables. Our simulations show a peak testing accuracy of 66.5% and a peak training accuracy of 67.2% (Fig. 4a). This result confirms that our PT-ONN is as expressive as the MZI based ONNs. The confusion matrix is reported in Supplement 1, Section 2 (see Fig. S2). Furthermore, the training process was performed three times and the standard deviation is reported in Supplement 1, Section 3 (see Figs. S3 and S4).

To further assess the robustness of our PT-coupler based architecture, we simulate the PT-ONN under a noisy environment. Since the gain factors are used as training parameters, we consider their variation as the main source of error. For this study, the gain contrast dependent parameters $\theta s$ are perturbed by a gaussian distribution $p(\Delta\theta_i) = \exp(-\Delta\theta_i^2/(2\sigma^2))/\sqrt{2\pi\sigma}$, where $\sigma$ represents the strength of the noise. These perturbed 2-level systems will result in new transfer functions for the network. Under this scenario, we use the same technique to simulate on-chip training and report the influence of noise level on the final training and testing accuracies in Fig. 4b. As compared to the result reported in [8], where the network has significant performance degradation when $\sigma$ exceeds 0.01, our design appears to be more resilient to noise.

## 4. Discussion

In this work, we demonstrated that PT-ONN architecture using gain-loss contrast as the training parameter can achieve comparable on-chip training and testing accuracies to that reported in the ONNs composed of MZI devices with phase shifters. Our PT-ONN also shows robustness to variations of its parameters ($\theta's$), in addition to having the advantages of smaller footprint, lower power consumption, and perhaps higher training speed.

In our implementation of the PT-ONN, $|\theta's|$ remain below 0.2. The distribution of the gain-loss contrast parameters ($\theta's$) is shown in Fig. 5, where most coefficients happen to be in the $-0.1$ to $0.1$ range and the average $\theta$ value is approximately $-2 \times 10^{-4}$. Our electromagnetic simulations show that a low to moderate level of gain will be adequate to reach the desired network performance. If the length of the coupling region is selected to be $z = 25\ \mu m$, one can adjust the spacing between the two waveguides to tune the strength of the coupling coefficient ($\kappa$) in order to keep the required gain within the attainable range afforded by III-V semiconductor materials. For example, for a coupler operating at a wavelength of $1.55\ \mu m$, and a coupling coefficient of $\kappa = 4 \times 10^4\ m^{-1}$, the maximum required gain coefficient is $\gamma = g/2 = 80\ cm^{-1}$, and the average gain per coupler is $\gamma = g/2 = 20\ cm^{-1}$ (given that average value of $|\theta's|$ is 0.05), which are well within the attainable range in most InGaAsP quantum wells structures. One should notice that the length may be further reduced by choosing $Z$ to be smaller than unity.

We also compared our PT-ONN against the MZI-based network in terms of footprint, switching speed, and power consumption. Because they share the same network topology, we only compare individual PT and MZI blocks. The state-of-the-art Joule heaters are reported to have a $\pi$ phase shift with a power requirement on the order of $20\ mW$ and a switching time of a few microseconds, with the reported length of the heater to be a few hundreds of micrometers [43]. On the other hand, for the maximum gain of $80\ cm^{-1}$, a PT-coupler at a length of $25\ \mu m$ requires $\sim 220\ \mu W$ of power to amplify a $1\ mW$ signal. However, the average power required per PT-coupler is merely $\sim 50\ \mu W$. Even at a quantum efficiency of 10%, the required power is $\sim 0.5\ mW$, which is still considerably lower than what is reported for phase shifters. Semiconductor amplifiers can also be modulated at a sub-nanosecond time scale [44].

One additional benefit of this approach is the possibility of implementing the entire PT-ONN using III-V semiconductor materials in a monolithic fashion. The required gain/loss can be achieved by pumping, and one possible candidate for realizing nonlinearity is III-V saturable absorbers [3,17]. Waveguides can be realized using QWI (quantum well intermixing) method [45–47] which changes the refractive index of III-V materials through inducing defects, or selective area regrowth. Finally, the detectors can be implemented on chip through an epitaxial regrowth process. With the advancements in heterogenous integration, one can also envision a multi-material platform to achieve the desired functionalities.

While in this study, we remained faithful to the exact PT-symmetric coupler, it is well known that the functionality of this device remains primarily unaffected if one of the waveguides is nominally loss free (e.g., through intermixing) and gain/loss is applied exclusively to the other waveguide. Novel designs for PT-couplers that allow more fabrication-friendly arrangements can be explored in the future works.

It should be noted that compared to MZI-based ONNs (like that of [3]), our PT-ONN cannot easily map an existing weight matrix onto the hardware by algorithmically computing the corresponding on-chip parameters. This is also the case for some quantum neural networks [48]. While this mapping will hardly lead to a functioning platforms due to hardware variances (or component imprecision), it nevertheless provides a good starting point from which one can fine-tune the network using on-chip training methods [18,49]. It may be of future interest to find better strategies to initialize the on-chip parameters of PT-ONNs.

Varying gain across the array may seem advantageous when compared to changing phases, in terms of time, power, and space management, however, it also introduces extra noise due to spontaneous emission. In the gain region, electrons in the excited state could spontaneously drop to a lower state and emit photons that are not necessarily coherent with respect to the incoming signal. Although the above simulation accounts for random gain variations, further analysis may be needed to more quantitatively assess the role of spontaneous emission noise in PT-ONN architectures. In addition, phase-intensity coupling can further complicate the training mechanism by introducing nonlinearity in the couplers. However, our current system is not expected to be severely affected by this effect due to the low average gain/loss contrasts. This aspect is discussed in Supplement 1, Section 4. Nonetheless, as the network grows this could become an issue that needs further consideration.

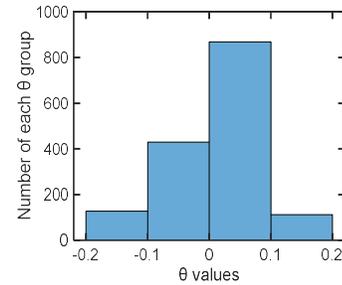

**Fig. 5**. Distribution of gain-loss contrast variable parameters ($\theta's$) in PT-ONN.

In conclusion, in this work we introduced for the first time an expressive III-V network based on PT-symmetric couplers for implementing reconfigurable ONNs without requiring changing phases. Our work may open up new avenues for realizing fast, efficient, monolithic, and compact optical neural networks on chip.

**Funding.** Air Force Office of Scientific Research (FA9550-20-1-0322, FA9550-21-1-0202), DARPA (D18AP00058), Office of Naval Research (N00014-19-1-2052, N00014-20-1-2522, N00014-20-1-2789), Army Research Office (W911NF-17-1-0481), National Science Foundation (ECCS CBET 1805200, ECCS 2000538, ECCS 2011171), and US–Israel Binational Science Foundation (BSF; 2016381).

**Acknowledgment**. The authors acknowledge fruitful discussions with Demetrios Christodoulides from CREOL, UCF, and Jiaqi Gu from Texas A&M University. The authors also appreciate the help from Omid Hemmatyar, Yuzhou Liu and Andrew Wilkey for technical support.

**Disclosures.** The authors declare no conflicts of interest.

**Data availability.** Data underlying the results presented in this paper are not publicly available at this time but may be obtained from the authors upon reasonable request.

**Supplement document**. See Supplement 1 for supporting content.